\documentclass[twocolumn,showpacs,preprintnumbers,amsmath,amssymb]{revtex4}

\usepackage{graphicx}
\usepackage{bm}

\begin{document}

\preprint{APS/123-QED}

\title{Osmotically Driven Shape Transformations in Axons}

\author{Pramod A. Pullarkat$^{1}$, Paul  Dommersnes$^{2}$, 
Pablo Fern\'andez$^{1}$,
Jean-Fran\c{c}ois Joanny$^{2}$, and Albrecht Ott$^{1}$}

\affiliation{$^{1}$Experimentalphysik I, University of Bayreuth,
D-95440, Bayreuth, Germany \\
$^{2}$Institut Curie, UMR 168, 26 rue d'Ulm, F-75248,
Paris Cedex 05, France}

\date{\today}

\begin{abstract}
We report a cylindrical--peristaltic shape transformation in
axons exposed to a controlled osmotic perturbation. The peristaltic
shape relaxes and the axon recovers its original geometry within
minutes. 
We show that the shape instability
depends critically on swelling rate and  that volume and membrane area
regulation are responsible for the shape relaxation.  We propose that
volume regulation occurs via leakage of ions driven by elastic
pressure, and analyse the peristaltic shape dynamics
taking into account the internal structure of the axon.  The
results obtained provide a framework for understanding peristaltic shape
dynamics in nerve fibers occurring {\it in vivo}.
\end{abstract}

\maketitle

Axons and dendrites, collectively known as neurites, are thin, tubular
extensions produced by neuronal cells \cite{alberts}. Structurally, they 
consist of an outer lipid membrane sheath bound to a core made up of an
elastic network of highly cross-linked biopolymers known as the cytoskeleton.
Neurites are known to lose their normal cylindrical geometry
and become peristaltically modulated--a process commonly known as
beading--under a wide range of situations.  These include
neuro-degenerative diseases like Alzheimer's \cite{alzheimer}, brain
trauma \cite{trauma}, stretch injuries to nerves \cite{ochs} and 
{\it in vitro} as well as {\it in vivo} application of neurotoxins
or drugs \cite{kainic-acid_curvature}, among others. 
In stretch injuries, tension in the axon is responsible 
for beading. In the other examples, disruption of cytoskeletal integrity
appears to be a common feature. 

Peristaltic modes have been studied in tubular membranes
and gels under tension \cite{bar-ziv1, nelson, ken}.  This 
``pearling instability'' is driven by surface tension, as in the 
Rayleigh-Plateau instability of liquid columns \cite{chandra}. In liquid
columns, peristaltic perturbations with wavelengths larger than the
circumference grow as they reduce surface area at constant volume. 
Unlike liquid columns, cylindrical vesicles/gels
are unstable only beyond a critical tension, when the gain in surface
energy due to a reduction in area overcomes the elastic energy for
deforming the membrane/gel. Similar arguments can explain peristaltic
modes observed in stretched nerves \cite{ochs-model} and cell
protrusions exposed to toxins \cite{bar-ziv3}. In all the above mentioned
biological examples the beaded state persists and no recovery has been
observed.

\begin{figure}
\includegraphics[width=6.5cm]{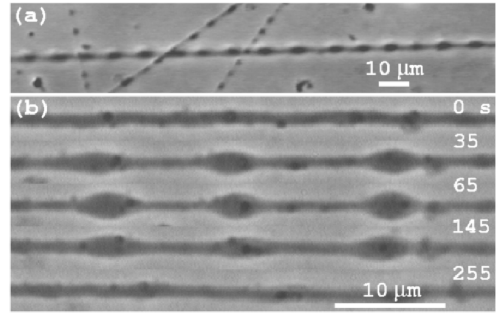}
\vspace{-0.3cm}
\caption{\label{fig:pearl}
(a) Osmotically induced shape instability in chick-embryo neurons.
(b) Image sequence showing the growth and relaxation
of the instability in a PC12 neurite.
}
\vspace{-0.5cm}
\end{figure}

We report a novel, reversible, peristaltic instability observed in
chick-embryo axons and in PC12 neurites \cite{culture}
after a sudden dilution of the external medium, as shown
in Fig.\ \ref{fig:pearl}.  The transition to the peristaltic state
occurs at a threshold membrane tension, induced by osmotic
swelling. However, unlike previous reports on beading,
the osmotically induced
instability relaxes and the neurite recovers its original shape within
minutes. We show that the relaxation is due both to cell volume and
membrane surface regulation. Remarkably,
the shape stability depends critically on the swelling rate.
We model the pearling/beading instability taking into account the gel-like
dynamics of the cytoskeleton and the membrane-cytoskeleton coupling,
and find wavelengths in agreement with
experimental observations. Further, we propose that elastic stresses
in the cytoskeleton provide the driving force for water and ions to 
leak through channel proteins in the membrane \cite{alberts} 
to effect volume regulation.
%
%
%

{\bf Experiments.-} 
The experiments are performed using a $10\times5\times1$ mm$^3$ 
flow chamber with temperature maintained within
$\pm0.2$ $^\circ$C. The cells are grown on treated glass coverslips
\cite{culture} and transfered to the flow-chamber. The flow velocity in 
the plane of the axon is about 1 $\mu {\rm m s}^{-1}$ and
the shear rate is about 1 ${\rm s}^{-1}$, well below values needed to 
cause any significant shear stress on the neurites. The flow is switched 
from normal medium, 
with an ion concentration $\phi_0 = 0.3$ mol/l, to a diluted medium 
to induce a hyposmotic shock. Phase contrast images are recorded
using a CCD camera at a magnification of 0.084 $\mu$m/pixel. The
neurite boundaries are traced with a typical error of a couple of
pixels using a home-developed image analysis software. The volume and
area are computed assuming axial symmetry. Only neurites which are
firmly adherent at the two ends and freely suspended along the length 
are used for measurements.

\begin{figure}
\includegraphics[width=7.5cm]{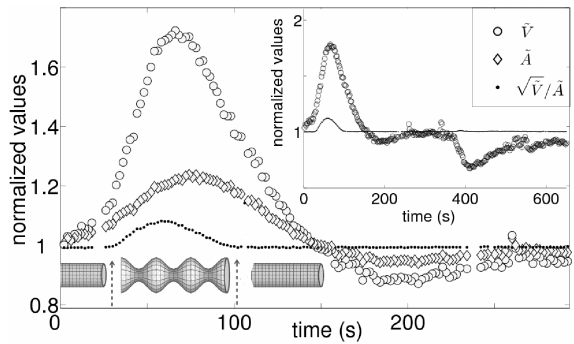}
\vspace{-0.5cm}
\caption{\label{fig:vol-area}
Evolution of the normalized volume $\tilde{V}$ and area 
$\tilde{A}$ of
a neurite with $R_{0}$ = 0.5 $\mu$m at 33 $^{\circ}$C. The ratio 
$\sqrt{\tilde{V}}/\tilde{A}$
(dotted line) reflects any deviations from the cylindrical geometry.
The inset shows the volume and shape responses when the 
concentration is switched in the sequence $\phi_0 \stackrel{0 {s}}
{\rightarrow} 0.5\phi_0 \stackrel{370 {\rm s}}{\longrightarrow} \phi_0$.
The initial slow swelling is due to a non-steplike change in
concentration, as revealed by using an absorbing dye.
Note: the shape remains cylindrical for the $0.5\phi_0 \rightarrow \phi_0$ 
shock. 
}
\vspace{-0.5cm}
\end{figure}

After exposure to a hyposmotic solution the neurite immediately
starts to swell with a uniformly increasing radius from its initial
value $R_{0}$. Fig.\ \ref{fig:vol-area} shows the typical variation of
the normalized volume $\tilde{V} = V/V_{0}$ and area $\tilde{A} = A/A_{0}$.
For sufficiently large osmotic shocks, a standing peristaltic mode sets 
in at a threshold radius $R_{c}$. The plot of $\sqrt{\tilde{V}}/\tilde{A}$ 
shows the evolution of the peristaltic shape. An increase in 
$\sqrt{\tilde{V}}/\tilde{A}$ indicates a reduction in surface area 
compared to a cylinder with identical volume.  The mode amplitude 
increases with time, reaches a maximum and then decreases back to zero. 
The amplitude and the wavelength appear uniform along the entire length 
of the neurite (see Fig.\ \ref{fig:pearl}).  An analysis of the instability 
observed on more than a hundred PC12 neurites with lengths of 
$\sim 50-500$ $\mu$m reveals the following.
%
%
\newline (i) 
The instability occurs only above a critical dilution, which is about
$0.5\phi_0$ at 37 $^{\circ}$C and about $0.7\phi_0$ at 25 $^{\circ}$C
for an initial radius $R_{0} \simeq 0.7$ $\mu$m.  The critical
dilution  decreases for smaller $R_{0}$ or for lower temperature.
\newline (ii)
The wavelength $\lambda$ increases with increasing $R_{0}$ but is
independent of the neurite length. For a given neurite, the
wavelength remains approximately constant during both the growth and
relaxation of the instability.
\newline (iii)
The maximum volume $V_{m}$ scales linearly with $V_{0}$, i.e.,\
$V_{m}/V_{0}$ is largely independent of the neurite selected.
\newline(iv)
Typically, the instability sets in when the area increases by 3--8\%.
The maximum increase can reach 20\%.
\newline (v)
For osmotic shocks slightly below the critical value, no shape change
occurs even when the area increases well beyond the 3--8\% mentioned
earlier. The growth rate of the area decreases
with decreasing osmotic shock.  Therefore, we conclude that the
stability depends on the rate of area increase as well as its
magnitude. In another set of experiments, when the dilution is
performed in several steps of very weak osmotic shocks 
the neurites remain cylindrical even when the external
solution is diluted to pure water. In this case the volume and area
remain close to their initial values throughout the experiment.
\newline (vi)
Water flow across the membrane is mainly through water specific channel
proteins belonging to the aquaporin family \cite{agre_review}. Blocking 
them using 500mM dimethylsulfoxide significantly reduced swelling and made the neurites more resistant to peristaltic shape change.

{\bf Discussion.-}
{\it Volume regulation.---} 
We assume the neurite to be a cylindrical, isotropic
gel with an adhering fluid membrane on its surface.
Based on fluorescence microscopy, we conclude that the membrane is firmly attached to the cytoskeleton at least during the onset of the instability.
The osmotic pressure difference across the membrane is $\Delta
\Pi = kT(\phi_{i}-\phi_{e})$, where $\phi_{e}$ and
$\phi_{i}$ are the external and internal net ion concentrations. The
volume flux of water through the membrane is
\begin{equation}
\label{eq:water}
J_{w}=L_{p}
\left(\Delta \Pi-\Delta p\right),
\end{equation}
where $L_{p}$ is the membrane permeability to water and $\Delta p$
is the hydrodynamic pressure difference over the membrane. The gel
response time is faster than the timescale of swelling, which is
determined by $L_{p}$.  The local normal force balance at the
membrane determines the hydrodynamic pressure jump $\Delta
p=(K+4\mu/3) \Delta R/R_0+E\Delta R/R_0^2 =\tilde{K}\Delta R/R_0$,
where $R=R_0+\Delta R$ is the neurite radius, $K$ and $\mu$ are the
compression and shear moduli, and $E$ the stretching modulus of the
membrane. The volume growth rate of the neurite is 
$\dot{V} =J_{w} A$.

Volume relaxation occurs due to the opening of membrane-bound
ion-channel proteins \cite{alberts}.  This Regulatory Volume Decrease
is known to involve both ``active'' as well as ``passive'' processes, many
of which are poorly understood and cell-type dependent \cite{volreg_rev}.
In general, it is known that the ion conductivity $G$ depends of the charge
distribution across the membrane. However,
it has been shown for PC12 cells that the membrane potential vanishes during swelling induced by an hyposmotic shock and it does not
recover during the entire relaxation phase \cite{memb_pot, volreg_rev}. Hence, we assume a constant average $G$ 
for simplicity and write the ion flux as 
\begin{equation}
\label{eq:ion}
J_{\rm ion} = GkT \ln{\frac{\phi_{e}}{\phi_{i}}}.
\end{equation}
The time variation of the internal number $N$ of ions is $\dot{N} =
J_{\rm ion} A$. Rescaling the conservation
Eqs.~(\ref{eq:water},\ref{eq:ion}) reveals two characteristic
timescales: $\tau_{w}=\frac{R_0}{2 L_{p} \Pi^0 } $ for water influx
and $\tau_{\rm ion}=\frac{\Pi^0 R_0}{2G(kT)^2}$ for ion conduction,
where $\Pi^0 \approx 700$ kPa is the osmotic pressure of normal
medium.  For weak osmotic shocks the conservation
Eqs.~(\ref{eq:water},\ref{eq:ion}) can be linearized, resulting in a
double exponential relaxation form for the volume, with characteristic
timescales related to $\tau_w$ and $\tau_{\rm ion}$.
At short times the volume change is given by 
$\Delta V=L_{p}\Delta \Pi_{e} A_0 t$, where 
$\Delta \Pi_{e}=\Delta \Pi(t = 0)$. Thus, the initial swelling is
not influenced by elastic forces or ion flow. 
$L_{p}$ measured from experiments increases
from $L_{p} = 0.8-1.6(\pm 0.3)\times 10^{-14}$ m/(Pa.s), 
in a 15--37 $^{\circ}$C range.  

\begin{figure}
\begin{tabular}{@{}lr@{}}
\includegraphics[width=3.6cm]{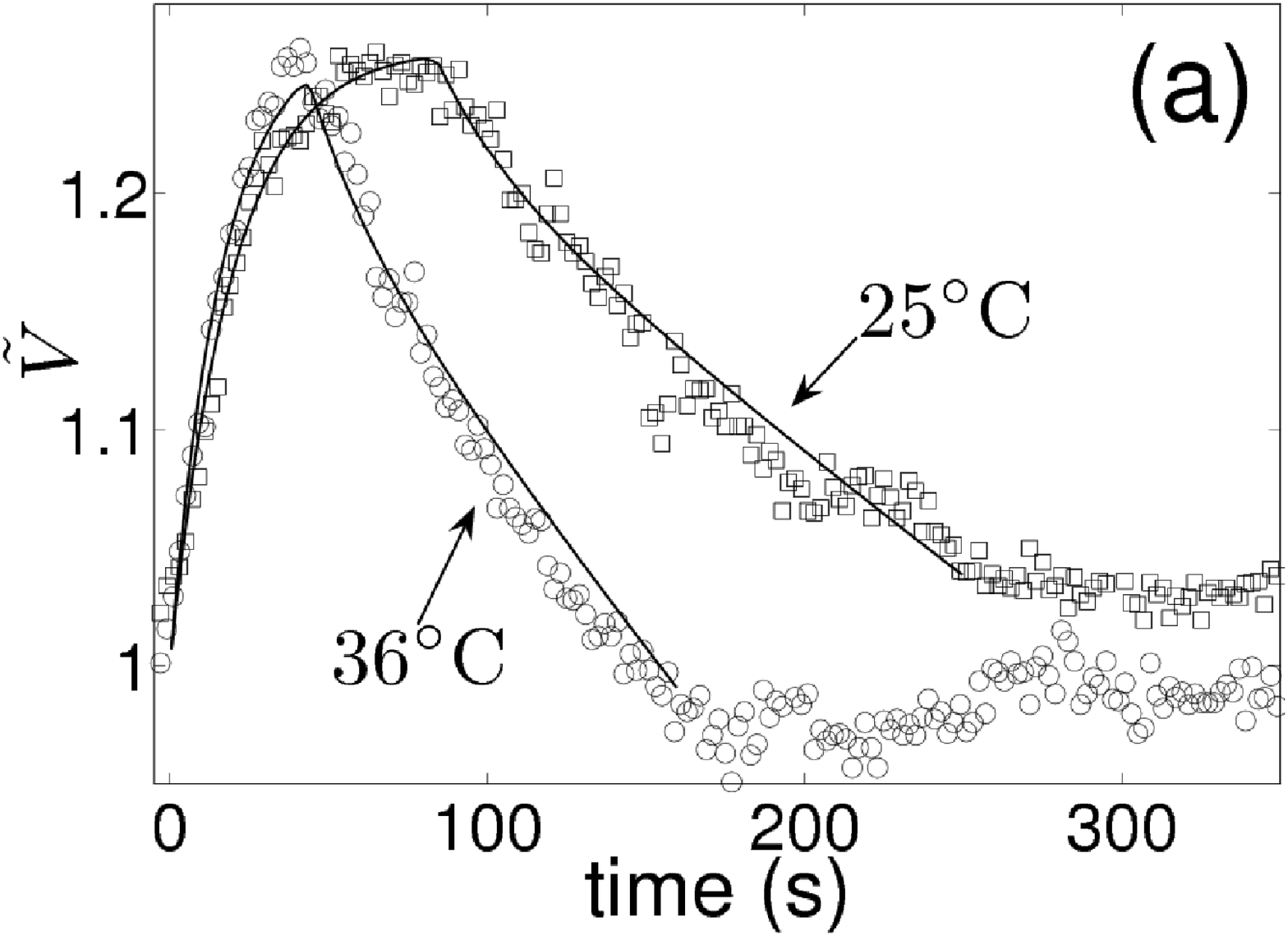}
\includegraphics[width=3.6cm]{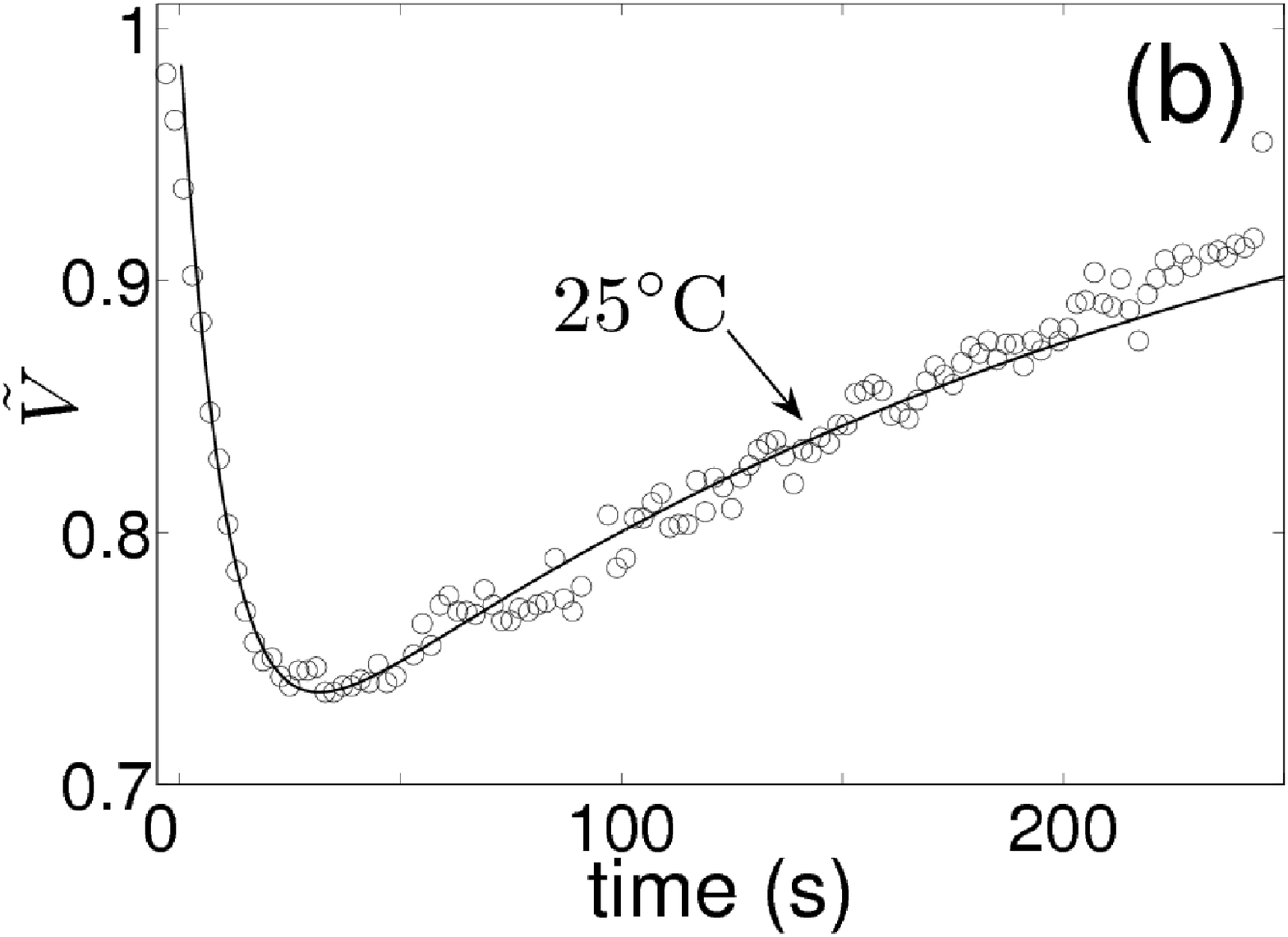}
\end{tabular}
\vspace{-0.3cm}
\caption{\label{fig:relax-model}
Comparison between volume evolutions obtained experimentally and from
the analysis. (a) For $\phi_0 \rightarrow 0.7\phi_0$ shock. 
$\tilde{K} = 0.025\Pi^{0}$,
$P_{a} = 0.06\Pi^{0}$, $\tau_w = 22$ s, and $\tau_{\rm ion} = 70$ s for
$25^{\circ}{\rm C}$.
$\tilde{K} = 0.03\Pi^{0}$, $P_{a} = 0.09\Pi^{0}$, $\tau_w = 18$ s, and 
$\tau_{\rm ion} = 60$ s for $36^{\circ}{\rm C}$.
(b) For $0.5\phi_0 \rightarrow \phi_0$, using $\tilde{K} = 1.5\Pi^{0}$, 
$P_{a} = 0$, $\tau_w = 34$ s, and $\tau_{\rm ion} = 24$.  
The neurite is densely filled with cytoskeleton and organelles
which have to be compressed close to their ``dead volume'' for $V<V_{0}$
giving a higher effective value of $K$ for (b).
}
\vspace{-0.6cm}
\end{figure}

Numerical analysis of the volume response given by
Eqs.~(\ref{eq:water},\ref{eq:ion}) produces either a smaller swelling
or a longer relaxation as compared to experiments. To account for
these, we consider the following biological processes. (i)
The open probability of mechano-sensitive ion channels increases with
membrane tension \cite{alberts}. Therefore, we expect an increase in
ion conductivity upon swelling. Including a tension dependence on $G$,
however, cannot account for the fast relaxation and large swelling.
(ii) Filamentous actin undergoes a transient depolymerization 
mainly due to an influx of $\rm{Ca}^{2+}$ followed by repolymerization 
and contractility, due to myosin activation by $\rm{Ca}^{2+}$, during
volume relaxation \cite{volreg_rev, active}. Thus, this actin
dynamics may account for both increased swelling and fast relaxation.
Indeed, including an active pressure $P_{a}$ in Eq. \ref{eq:water},
after the maximum in volume, gives quantitative agreement with
experiments as in Fig.\ \ref{fig:relax-model}(a). From the fits, we
estimate $G=0.7-4\times 10^{38}/({\rm J m^2 s})$. Converting to
electrical conductivity units gives $G\equiv 2-10\; {\rm S/m^2}$,
in agreement with reported values \cite{weiss}. We also estimate
$P_{a} \sim 0.05-0.1\Pi^0$.  At sufficiently high
temperatures, we often observe an ``undershoot'' of the volume (see
Fig.\ \ref{fig:vol-area}), which could be attributed to active
contraction of actin.  This complex actin dynamics has not been
reported for hyperosmotic shocks (sudden increase in external
concentration).  Indeed, the simpler analysis without active pressure
fits well to our hyperosmotic shock data as shown in Fig.\ 
\ref{fig:relax-model}(b).  Shrinking responses have previously been
analyzed for kidney cells in suspension assuming volume dependent
conductivities and neglecting elastic pressure \cite{brazil}.

%
%
%
%

{\it Shape instability.---}
As the neurite swells, the membrane tension $\sigma$ increases and for
a critical swelling radius $R_{c}$ it is unstable to peristaltic
shape modulations.  In \cite{bar-ziv3} a scaling argument was
developed to determine the fastest growing wavelength in cylindrical
cells, assuming that dissipation is due to viscous flow inside the
cell.  Here, we take into account the gel-like structure of the neurite
cytoskeleton. The solvent flow inside the gel is given by
$
-\nabla p+\eta\nabla^2{\bf v}-\zeta({\bf v}-\partial_t {\bf u})=0, 
$
where $\bf v$ is the solvent velocity ($\nabla\cdot {\bf v}=0$), $p$
the hydrodynamic pressure, $\bf u$ the gel elastic deformation, $\eta$ the
solvent viscosity and $\zeta$ the gel-solvent friction. In addition
the gel elastic force is balanced by hydrodynamic friction
$
\zeta (\partial_t{\bf u}-{\bf v})=\mu \nabla^2 {\bf u} 
+(K+\frac{1}{3}\mu) \nabla(\nabla\cdot {\bf u}).
$
The solvent flow outside the neurite is given by the Stokes equations.  We
consider the stability of a modulation in the neurite radius 
$R=R_{c}+\epsilon e^{\omega t} \sin(q z)$,  
with amplitude $\epsilon$ and growth rate $\omega(q)$.  
We  assume that the gel is in elastic equilibrium at the onset of
the instability, i.e. radial compression modes have relaxed. 

The membrane imposes boundary conditions at the neurite surface. The
solvent velocity is continuous, and the solvent flow through the
membrane is governed locally by Eq.~(\ref{eq:water}). The osmotic pressure
difference is preserved within linear instability analysis, since a 
peristaltic modulation conserves volume to linear order. Conservation of 
membrane area implies $\partial_t u_r+R_{c}\partial_z v_z=0$ \cite{nelson}.  
The total radial stress jump (hydrodynamic and elastic) at the neurite
surface must balance the Laplace pressure due to membrane tension
$\Delta\sigma_{rr}= P_{\rm Laplace}$.  Tangential force balance
implies $\sigma^{E}_{rz}=\xi (v_z-\partial_t u_z)$, where $\sigma^{E}$ 
is the gel elastic surface stress, and $\xi$ the sliding
friction coefficient between the cortical actin network and membrane.
Rescaling the equations reveals two timescales for the gel dynamics:
the relaxation time of compression modes, $\tau_C=\zeta R^2/K$, and
that of shear modes $\tau_S=\eta/\mu$.  The compression modes relax
slower than the shear modes, $\tau_C \propto (R/\ell)^2 \tau_S$, where
$\ell$ is the mesh size of the gel. For a shear modulus $\mu \sim 1$ kPa 
and solvent viscosity 50 times water viscosity, $\tau_{S}\sim
10^{-4}$ s.

We find that the membrane-cytoskeleton sliding friction $\xi$ affects
the instability to order $\xi/(\zeta R)$. Assuming
that the friction is due to anchored membrane proteins, and using
common values for membrane viscosity, we estimate that $\xi \ll \zeta
R$, showing that $\xi$ can  be neglected. 

\begin{figure}
\begin{tabular}{@{}lr@{}}
\includegraphics[width=3.5cm]{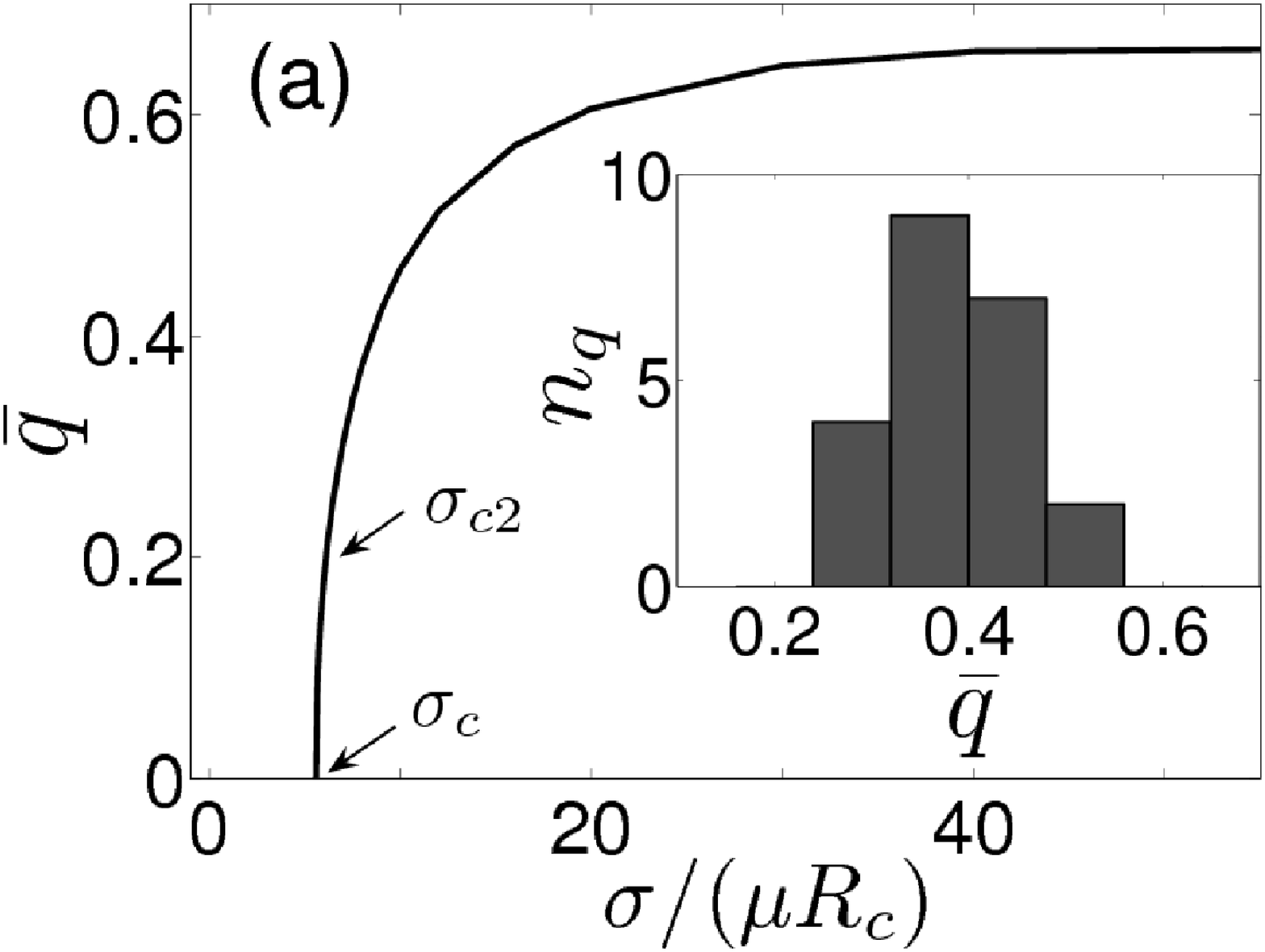}
\includegraphics[width=3.5cm]{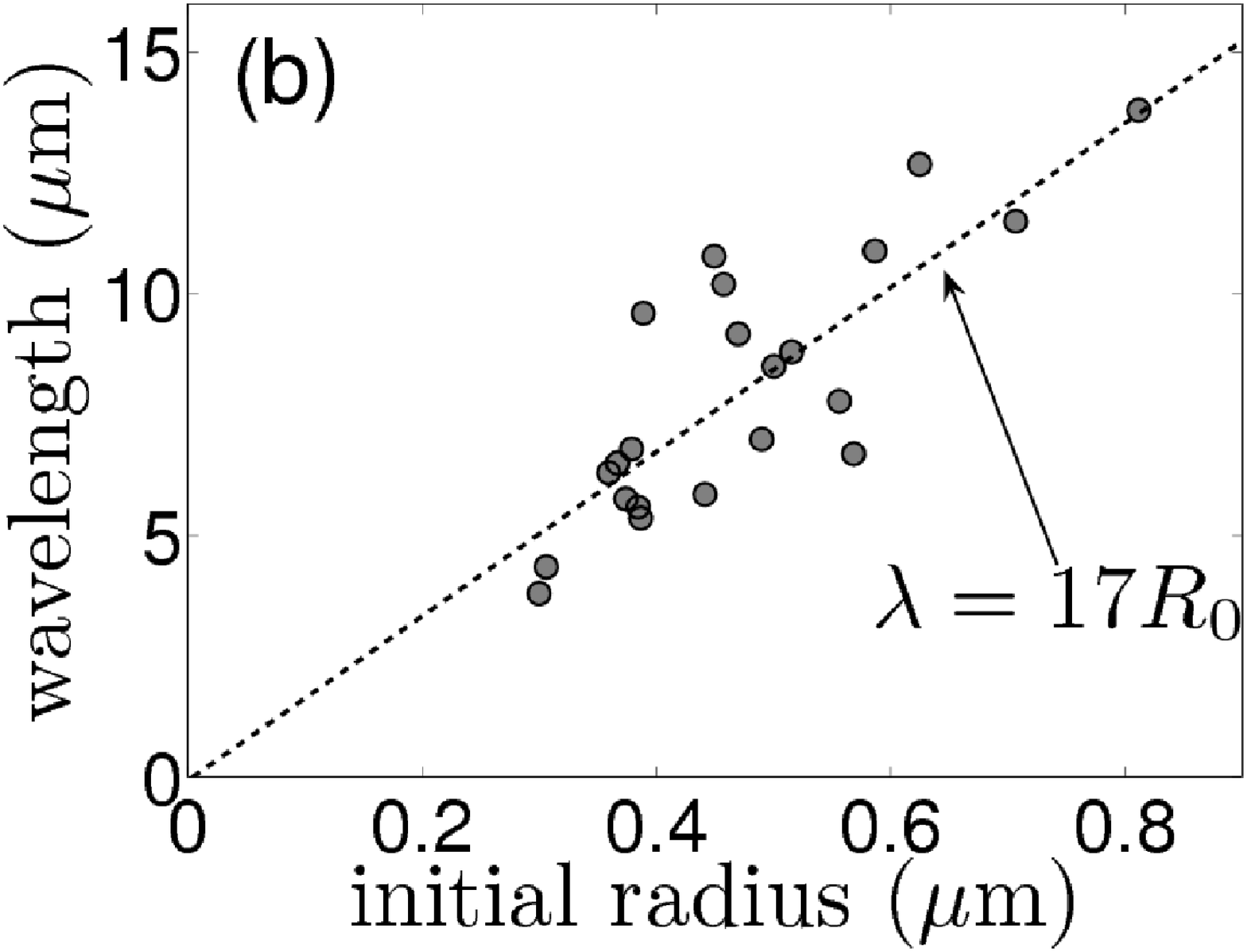}
\end{tabular}
\vspace{-0.4cm}
\caption{\label{fig:lambda}
(a) Theoretical plot for the fastest growing wavenumber with tension
and histogram of observed wavenumbers. 
(b) Variation of $\lambda$ with $R_{0}$ at 25 $^{\circ}$C for
$\phi_0 \rightarrow 0.67\phi_0$.
The scatter in the data is due to natural variations between cells.
}
\vspace{-0.5cm}
\end{figure}

The neurite is unstable above a critical tension 
$\sigma_{c}=\frac{6 K \mu R_c}{K+4/3 \mu} +{\cal O}({\bar{q}}^4)$. 
Close to $\sigma_c$ the fastest growing wavenumber is 
$\bar{q}= q R_{c}=\sqrt{\frac{\sigma-\sigma_{c}} {2\sigma_{c}}}$ with a
corresponding growth rate $\omega \sim \frac{1}{\tau_{C}}
(\frac{\sigma-\sigma_{c}} {\sigma_{c}})^2$, thus initially the 
dynamics is governed by the gel compression modes,
which is qualitatively similar to the result in \cite{ken}. Above a
second critical tension $\sigma_{c2}=6\mu R_c$, the instability grows
with the much faster rate $\omega\sim 1/\sqrt{\tau_{S}\tau_{C}}$. The
tension $\sigma_{c2}$ corresponds to the critical tension for pearling
instability in an incompressible gel ($\sigma_c \rightarrow
\sigma_{c2}$ when $K\rightarrow \infty$), which suggests that above
$\sigma_{c2}$ the instability grows via a peristaltic shear
deformation of the gel.  For high tension the fastest growing mode
converges to the universal value $\bar{q}\approx 0.65$, independently
of the gel elasticity and gel-solvent friction [see Fig.\ 
\ref{fig:lambda}(a)].

Using $K \sim 20$ kPa, estimated from volume relaxation analysis,
and $\mu \sim 1$ kPa \cite{fabry_shear} gives 
$\sigma_{c2}\approx 1.07 \sigma_c$ and
$\bar{q}_{c2}\approx 0.2$. For $R_0 \sim 0.5$ $\mu$m the
critical tension for pearling is $\sigma_{c}\sim$ $3 \times 10^{-4}$ N/m.

The above analysis sets an upper limit $\bar{q}<0.65$ on the fastest growing
wavenumber which is in good agreement with
observations [see Figs.\ \ref{fig:lambda}(a,b)]. Remarkably, we never
observed wavenumbers below $\bar{q} \lesssim 0.2$, even for dilutions close
to the pearling threshold. This could be due to the steep increase in
$\bar{q}$ close to $\sigma_c$. Also, the timescale for the growth of modes 
with small $\bar{q}$, which occur close to $\sigma_c$, may be longer than the
volume regulation time and hence these modes are not easily observable.

%

{\it Surface area regulation.---}
Figure \ref{fig:vol-area} shows some remarkable features: (i) the
instability begins to relax before the area and hence the apparent
tension has reached its maximum value, (ii) the instability
disappears when the measured area and volume
are well above their values at its onset and (iii) a 20\%
increase in area due to stretching alone
is unexpected as most biological membranes are known to rupture beyond
3--5\% of stretching. A reasonable explanation for these observations
is that lipids are continuously added to the membrane thereby relaxing 
tension. Assuming that the membrane tension is the same at the onset and 
disappearance of the instability, we estimate a rate of increase of area of
about $4 \times 10^{-3}$ $\mu$m$^2$/s per unit length. This process would lead
to a viscoelastic-like rate dependence of tension, giving rise to the critical
rate of dilution mentioned earlier. The increase in area is most likely due to
the fusion of tiny multi-lamellar vesicles to the outer membrane \cite{fusion0,
fusion1}. It is unlikely that this tension relaxation is due to flow of membrane
from the extremities as (a) the onset and relaxation of the instability appears
uniform along the length even in the longest neurites ($\sim 1$ mm) and (b) we
do not observe any flow of extraneous particles sticking to the outer membrane.

{\bf Conclusion.-}
Our experiments on the dynamics of the pearling instability in
neurites  reveal the role of volume regulatory
mechanisms and direct membrane tension regulation in maintaining or
recovering their normal cylindrical shape. 
Recovery to cylindrical shape has not been observed in previous
experiments on beading cells. A possible explanation is that osmotic
swelling is a gentle perturbation to which the cell has time
to react before the deformation results in irreversible
damage. Indeed, we observe that the instability is
irreversible for sufficiently strong osmotic shocks. 
Moreover, it is not just the magnitude of
osmotic swelling which triggers the instability, but also the rate
at which the cell swells. Thus, observation of the instability is
a direct way to probe the membrane surface regulation
dynamics in neurites.  We have also performed
preliminary experiments on pulling neurites using a glass needle.
There is a critical rate as well as magnitude of pulling
beyond which peristaltic modes appear. For small
amplitude modulations, the instability relaxes, 
whereas for strong modulations it persists.
Physiologically, the volume and membrane surface regulation
mechanisms should play an important role in stabilizing axon shape to
sufficiently slow and gentle perturbations occurring {\it in vivo}. The
theoretical analysis highlights the importance of cytoskeletal
elasticity in the volume regulation processes, and suggests that active
contractions of the cytoskeleton may play an important role.
The implications of our investigation to pathological conditions,
especially to dendritic beading due to local inflammation \cite{inflamation},
is an interesting topic for further research.

{\bf Acknowledgments.-}
We are extremely grateful to J. Prost for valuable discussions and suggestions. We thank the 
referees for interesting suggestions.
Preliminary experiments were conducted at Institut Curie, Paris, 
where P. A. P. was supported by Institut Curie and CNRS.
P. D. was supported by a Marie Curie fellowship.
Financial support from the European Commission is acknowledged under
project HPRN-CT-2002-00312. 

\bibliography{pearl1}

\end{document}